\documentclass[12pt]{iopart}
\usepackage{iopams}  
\usepackage{url}
\usepackage{cite}
\usepackage{lineno}
\begin{document}

\title[]{ Image denoising and model-independent parameterization for improving IVIM MRI}

\author{\textmd{Caleb Sample}\textsuperscript{\textmd{1,2}}, \textmd{Jonn Wu}\textsuperscript{\textmd{3,4}}, \textmd{Haley Clark}\textsuperscript{\textmd{1,2,4}}}
\vspace{0.5cm}
\address{\textsuperscript{1}Department of Physics and Astronomy, Faculty of Science, University of British Columbia, Vancouver, BC, CA\\
\textsuperscript{2}Department of Medical Physics, BC Cancer, Surrey, BC, CA\\
\textsuperscript{3}Department of Radiation Oncology, BC Cancer, Vancouver, BC, CA\\
\textsuperscript{4}Department of Surgery, Faculty of Medicine, University of British Columbia, Vancouver, BC, CA
}

\begin{abstract}
 Variability of IVIM parameters throughout the literature is a long-standing issue, and perfusion-related parameters are difficult to interpret. We demonstrate for improving the analysis of intravoxel incoherent motion imaging (IVIM) magnetic resonance (MR) images, using image denoising and a quantitative approach that does not require imposing specific exponential models. IVIM images were acquired for 13 head-and-neck patients prior to radiotherapy. Of these, 5 patients also had post-radiotherapy scans acquired. Image quality was improved prior to parameter fitting via denoising. For this, we employed neural blind deconvolution, a method of undertaking the ill-posed mathematical problem of blind deconvolution using neural networks. The signal decay curve was then quantified in terms of area under the curve ($AUC$) parameters. Denoised images were assessed in terms of blind image quality metrics, and correlations between their derived parameters in parotid glands with radiotherapy dose levels. We assessed the method's ability to recover artificial pseudokernels which had been applied to denoised images. $AUC$ parameters were compared with the apparent diffusion coefficient ($ADC$), biexponential, and triexponential model parameters, in terms of their correlations with dose, and their relative contributions to the total variance of the dataset, obtained through singular value decomposition. Image denoising resulted in improved blind image quality metrics, and higher correlations between IVIM parameters and dose. $AUC$ parameters were more correlated with dose than traditional IVIM parameters, and captured the highest proportion of the dataset's variance. V This method of describing the signal decay curve with model-independent parameters like the $AUC$, and preprocessing images with denoising techniques, shows potential for improving reproducibility and functional utility of IVIM imaging. 
\end{abstract}
\newpage 
\section{Introduction}
    Intravoxel Incoherent Motion (IVIM) Magnetic Resonance Imaging (MRI) is a diffusion imaging technique which quantifies the translational motion of molecules within imaging voxels \cite{lebihan_2019}. This translational motion is due to the random diffusion of water molecules within tissue, as well as perfusion effects. These translational motions are inferred by analyzing signal decay in each voxel as a function of applied diffusion gradient strengths (b-values). Le Bihan first introduced a method of quantifying these \textit{in vivo} motions using Stejskal and Tanner's \cite{Stejskal1965} diffusion gradient method in 1986 \cite{LeBihan1986}. This paper introduced the concept of an "apparent diffusion coefficient" (ADC). The pseudodiffusion coefficient quantifies the combined effects of real diffusion due to random molecular motion and the randomly oriented microcirculation of blood in capillary beds. By acquiring signals using multiple diffusion b-values, the ADC can be determined using the equation,
\begin{equation}
	S(b) / S(0) = e^{-ADC \cdot b}
\end{equation}
which is analagous to Stejkal and Tanner's \textit{in vitro} diffusion equation for a still liquid, $S(b) / S(0) = e^{-D \cdot b}$ \cite{Stejskal1965}, where D is the self-diffusion coefficient of the liquid. The ADC is a simple metric for describing the signal decay curve \textit{in vivo}; however, it does not model real diffusion, and is only a method of approximating the biological mechanisms responsible for the observed signal decay seen with increasing diffusion b-values. An important difference is that Stejskal and Tanner's diffusion equation comes from directly solving the Bloch equations \cite{bloch} for the magnetization vector, $\vec{M_0}$, as a function of time, while Le Bihan's application of this equation to \textit{in vivo} motion is an approximation based on this derivation. The most useful way to interpret the $ADC$ is to think of it as simply a parameter for approximating the \textit{in vivo} signal versus b-value curve as an exponential decay curve.
    
    It is theorised that contributions of diffusion and perfusion motion to the signal decay can be disentangled by fitting appropriate models to the signal decay curve in each voxel, as originally posed by Le Bihan in 1988 \cite{lebihan_1988}. This operates by modelling the collective perfusion of blood through microcapillaries as a "pseudodiffusion" process, with a pseudodiffusion coefficient, $D^*$. $D^*$ is typically on the order of 10 times larger than $D$ \cite{lebihan_2019}, and therefore, the signal decay is largely attributed to perfusion at low b-values, and diffusion at high b-values \cite{Zhou2016}. A common approach for separating diffusion and pseudodiffusion in practice is to apply a biexponential signal decay model to the data,
    \begin{equation}
        S(b)/S(0) = f \, e^{-b \cdot D^*} + (1-f) \, e^{-b \cdot D}
    \end{equation}
    where $f$ denotes the fraction of the signal decay attributed to pseudodiffusion within a voxel.
    
    Reproducibility of the pseudodiffusion coefficient remains a long-standing issue  while D and ADC have been shown to exhibit higher stability \cite{Koh_2006, Troelstra2022, Nai_2023, scalco_2023, GurneyChampion_2018, Liu_2021}. $f$ and $D^*$ have been shown to have inferior reproducibility to $D$; however, $f$ has been shown to be empirically useful for predicting age in nervous tissue \cite{Vieni_2020, Yamada_2023}.

    In recent years, the literature has seen a shift towards adoption of a triexponential model \cite{Kuai_2017, Cercueil_2014, Wurnig_2017, Chevallier_2019, Riexinger_2019, Riexinger_2020} of the form
    \begin{equation}
        S(b)/S(0) = f_1 \, e^{-b \cdot D_2^*} +  f_2 \, e^{-b \cdot D_2^*}+ (1-f_1-f_2) \, e^{-b \cdot D}
    \end{equation}
    Model-fit uncertainty can be reduced by introducing an extra exponential term \cite{Cercueil_2014, Wurnig_2017, Chevallier_2019, Riexinger_2019, Riexinger_2020}; however, it remains to be seen whether these parameters are reproducible or have practical physiological interpretations. 

    It is uncertain whether attempts to separate diffusion from perfusion are effective, as it has been reported that fitting a mono-exponential function, $S(b)/S(0) = (1-f) \, e^{-b \cdot D}$ is more reliable than a bi-exponential fit for differentiating pathological grades of esophageal squamous cell carcinoma (ESCC) \cite{Liu_2021}. Heightened measurement error of signal at low b-values \cite{LeBihan_1991, Pekar1992} leads to sub-optimal conditions for estimating perfusion effects, and estimation of $ADC$ in low b-value regions has shown poor reproducibility \cite{Koh_2009}. Noise-levels have also been shown to greatly impact parameter estimates \cite{Iima2015}.

    Poor reproducibility of perfusion-related parameters could be partially explained by over-simplification of physiological processes inherent in simple exponential models, as it has been shown that the optimal choice of model is dependent on tissue type \cite{Liao2021}. It is without doubt that, in reality, each imaging voxel containing biological tissue will encompass a complex arrangement of micro-structures, all having various perfusion fractions and molecular motion. Kuai et al \cite{Kuai_2017} used simulated data with 2-5 perfusion components to show that the variance of $f$ and $D^*$ tend to increase as the number of perfusion components increases, or the difference between pseudodiffusion components increases. 

    Reproducibility of IVIM parameters is also impacted by a lack of standardization for voxel sizes and b-value distributions, which several studies have attempted to optimize \cite{Perucho_2020, Lemke2011, Raju2021, Paganelli2023, Zhu2019, Riexinger2020}.
    Variation in voxel size is certain to impact multi-exponential model fits. For example, suppose there exists IVIM images with a voxel size of $2$ x $2$ x $2$ $mm^3$ and it is fitted with a bi-exponential model. That image is then immediately re-acquired with voxels of size $1$ x $1$ x $1$ $mm^3$ and again modeled with a bi-exponential fit. By comparing the two images, it is clear that the original image had been modelling octo-exponential components with a bi-exponential function. Averaging the parameters from the smaller voxels over the larger ones will, in general, yield different parameters than those of larger voxels exactly paired. Voxel-by-voxel analysis of IVIM images is difficult due to uncertainty, and parameters are therefore averaged over regions-of-interest (ROIs). This issue is exacerbated by partial volume effects (PVEs) from various tissue types and bleed-in from neighbouring voxels. 

    The goal of applying various signal-decay models to IVIM data is to describe the signal decay curve, which in turn, quantifies microscopic fluid motion within tissue. Bi- and tri-exponential parameters are particularly enticing as they appear to have a clear physiological interpretation. However, there is no consensus that $f$ is correlated with blood vessel density \cite{Iima2015, Lee2013, Bisdas2014}, and the ability of parameters to be interpreted physiologically is generally unimportant for practical applications. There is furthermore no consensus on the optimal IVIM model for describing the signal decay curve, with many alternative versions being recommended \cite{Kennan_1994, Wetscherek_2014, Henkelman1994, Duong2000, Fournet2016}. 
    
    IVIM parameters tend to be used to construct models for predicting various clinical end-points, without being directly interpreted. While model complexity has risen since the conception of IVIM MRI, the overall goal remains simple: to best describe the signal decay curve. We hypothesize that describing the signal decay curve without imposing any specific models onto the data will lead to higher reproducibility, and better modelling of outcomes.

    The purpose of this study is to introduce two methods for improving IVIM MRI efficacy and reproducibility. First, to improve image quality and reduce the effect of image aberrations on parameter estimates, we apply a denoising algorithm to each patient's diffusion images prior to model-fitting. Second, we introduce several model-independent parameters for quantifying the signal decay curve of IVIM MRI images, and compare them to bi-exponential, tri-exponential, and ADC parameters. 

\section{Methods}

    \subsection{2.1 Dataset}
        This study was approved by an institutional review board, and written, informed consent was obtained from all patients. The inclusion criteria for recruitment were: (1) a diagnosis of nasopharynx, base-of-tongue, or tonsil cancer to be treated with external beam radiotherapy; (2) expected to receive dose in parotid glands during treatment; (3) scheduled to have MR images acquired prior to radiotherapy and at 3 months following the last treatment day. IVIM MR images encompassing the parotid glands were acquired during routine clinical appointments for 12 head-and-neck cancer patients prior to radiotherapy (Age: 35-78Y, average: 60.4 ; sex: 11M, 1F). Of these, 5 had follow-up images acquired at 3 months post-RT (Age: 49-78Y, average: 59.3 ; sex: 5M, 1F). 4 of these 5 patients received weekly chemotherapy concurrent with radiotherapy (three patients: weekly 40 $mg / m^2$ cisplatin, one patient: weekly 20-40 $mg / m^2$ per week and 240 mg carboplatin).
    
        Images were acquired on a 1.5 T Magnetom Sola scanner (Siemens Healthineers) with an echo planar imaging sequence (EPI). Diffusion weighted images were acquired with a voxel size of 1.6 x 1.6 x (3.4-3.9) mm$^3$. This slight inter-patient variation in slice thickness was imposed to encompass parotid glands within 32 slices while maximizing the signal-to-noise ratio (SNR). The repetition time (TR) and echo time (TE) were 4900 ms and 105 ms, with a flip angle of 90$^\circ$. Images were acquired for 16 b-values (0, 20, 30, 40, 50, 60, 70, 80, 90, 100, 120, 150, 250, 400, 800, 1000). These values were chosen to follow recommendations for effectively separating perfusion parameters \cite{Cohen_2014, Ye2019}, typically with a partition margin around 50 $s/mm^2$ \cite{Raju2021}. Signals were acquired in 6 directions, spanning the 3 principal Cartesian axes and their vectors. Two signal averages were computed in each direction. Signals from various directions were combined by taking the geometric mean \cite{Mukherjee2008}.
    
        Parotid glands were manually contoured by a graduate physics student on clinical T1 images. These images were acquired with a turbo spin-echo (TSE) sequence (TR: 564 ms, TE: 20 ms, flip angle: 148$^\circ$). T1 images were acquired with a voxel size of 0.6 x 0.6 x 3 mm.
            
        DICOM radiotherapy dose and planning structure set files used for external beam radiotherapy treatment planning were exported from Varian Eclipse (Varian Medical Systems, Inc.) for computing dose statistics inside parotid glands. Whole-mean parotid gland dose levels were calculated within computed tomography (CT) contours, manually defined by a single, senior clinical radiation oncologist. 
        
    \subsection{2.2 Denoising}
        To mitigate the influence of partial volume and other blurring effects on IVIM parameter estimates, images were denoised using neural blind deconvolution. Blind deconvolution is the ill-posed mathematical problem of estimating hypothetical denoised images, $\vect{x}$, and their associated blur kernel, $\vect{k}$, which convolve to yield the original image, $\vect{y} = \vect{x} \ast \vect{k}$. This method is particularly useful in situations where spatial resolution and precision of signal localization is low, hence many studies have attempted to employ it for positron emission tomography (PET) \cite{kjell_2012, prev_1, prev_2, prev_3, prev_4}. Neural blind deconvolution is the method of solving this problem using neural networks which are simultaneously optimized to predict $\vect{x}$ and $\vect{k}$. It was originally implemented by Ren et al. \cite{ren_2020} in 2020 and applied to images from a 2-dimensional natural image database, and then significantly improved by Kotera et al. \cite{kotera_2021} in 2021. Neural blind deconvolution is an unsupervised learning methodology which trains and predicts on a case-by-case basis, without the requirement of a separate training set with ground truth images.

        Neural blind deconvolution was adapted for 3-dimensional prostate specific membrane antigen (PSMA) positron emission tomography (PET) in 2023, while incorporating simultaneous super-sampling into the methodology \cite{sample_2023_blind_deconv}. It was shown to improve blind image quality metrics, and strengthen correlations between PSMA PET uptake and sub-regional importance estimates in the parotid gland for predicting post-radiotherapy xerostomia (subjective dry mouth) \cite{sample_2023_psma_importance}. 
        
        Here we further build off of this methodology for suitability with IVIM MRI. First of all, we adapt the network to handle 4-dimensional input images (b, x, y, z), composed of 3-dimensional images for all b-values stacked. Images were normalized as $\vect{x} \rightarrow \frac{\vect{x} - \mu^{body}}{\sigma_{body}}$, where $\mu_{body}$ and $\sigma_{body}$ are the mean and standard deviation of signal in body voxels. The network architecture was maintained to predict a single kernel, such that a single blur kernel was predicted using all b-value images for each patient. 

        As the traditional formulation of the blind deconvolution problem, $\vect{y} = \vect{x} \ast \vect{k}$, accounts for bleeding of voxel signals, it does not account for background noise. As MRI signals contain rician noise \cite{Gudbjartsson1995}, we modified the mathematical formulation of the problem to include a noise term:
        \begin{equation}
            \vect{y} = \vect{x} \ast \vect{k} + \vect{\delta}
        \end{equation}
        
        This required the introduction of an additional symmetric convolutional auto-encoder network, $G_\delta$, for predicting the noise term $\vect{\delta}$. The networks used are thus,
        \begin{equation}
        G_x(\theta_x) = \vect{x} \,\,\,\, ,\,\,\,\, G_\delta(\theta_\delta) = \vect{\delta} \,\,\,\, ,\,\,\,\, G_k(\theta_k) = \vect{k} \,\,\,\, , \,\,\,\, \vect{y} = \vect{x} \ast \vect{k} + \vect{\delta}.
        \end{equation}
        where $\theta_x$, $\theta_\delta$ and $\theta_k$ represent trainable model parameters of $G_x$, $G_\delta$ and $G_k$, respectively. The following implicit constraints exist on network outputs which are satisfied automatically. 
        \begin{equation}
            0 \le G_x(\theta_x) \,\,,\,\, G_k(\theta_k) \ge 0 \,\,,\,\, \sum_i G_k(\theta_k)_i = 1.
        \end{equation}
        The optimization problem can be written in terms of the model parameters $\theta_x$, $\theta_\delta$ and $\theta_k$ as 
        \begin{equation}   
                \theta_x, \theta_k, \theta_\delta = \arg \min_{\theta_x, \theta_k, \theta_\delta} || G_x(\theta_x) \ast G_k(\theta_k) + G_\delta(\theta_\delta)- \vect{y}||_2^2 + R( G_x(\theta_x), G_k(\theta_k) ).
        \end{equation}

\begin{figure}[h]
          \centering
          \includegraphics[width=0.8\textwidth]{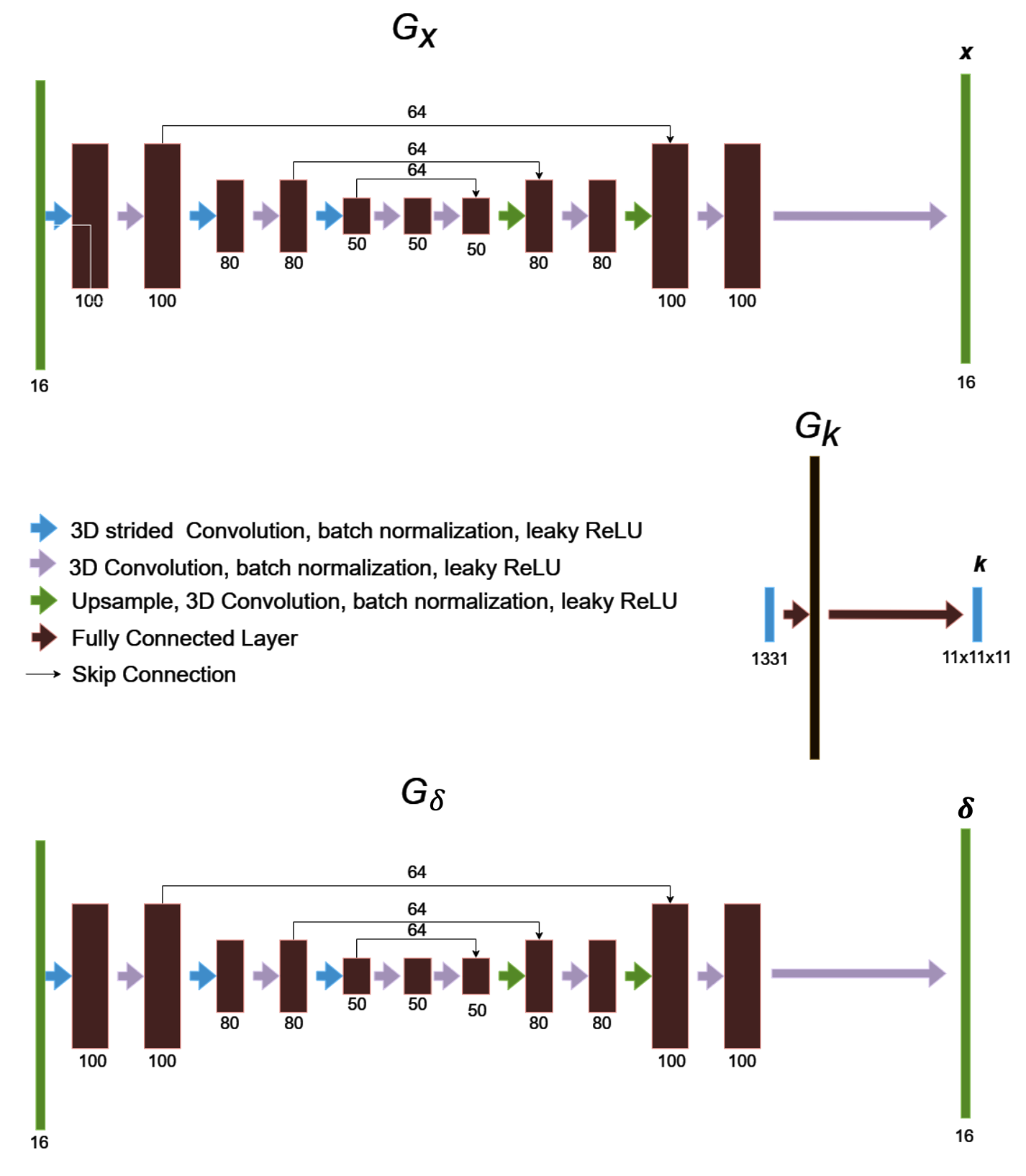}
          \caption{The blind deconvolution architecture used for denoising IVIM images is illustrated. $G_x$  and $G_\delta$ are symmetric, convolutional auto-encoder networks for predicting the denoised image, $\vect{x}$, and its additive noise contributions, $\vect{\delta}$. $G_k$ is a fully-connected network for predicting the blur kernel, $\vect{k}$. The fidelity loss term used in optimization compares the original image with $\vect{x} \ast \vect{k} + \vect{\delta}$ by using back-propagated model gradients from the loss function for iterative optimization.}
          \label{fig:architecture}
        \end{figure}

        The full architecture is summarized visually in Figure~\ref{fig:architecture}. 

        Optimization was performed using a single NVIDIA GeForce GTX 1060 GPU and a 2.8GHz Intel Core\textsuperscript{TM} i5-8400 CPU. As neural blind deconvolution is a computationally expensive procedure, further exacerbated by stacking multiple b-value images, images were cropped to extend at least 5 voxels outside of parotid gland borders in all 3 cartesian directions. This was performed separately for the left and right parotid glands, such that neural blind deconvolution was performed separately for each gland. Decreasing the image size in this manner was necessary to limit GPU memory usage.

        A multi-scale optimization procedure was implemented, as recommended for neural blind deconvolution by Kotera et al. \cite{kotera_2021} and recently implemented for PET denoising \cite{sample_2023_blind_deconv, sample_2023_psma_importance}. This involved pre-training networks using images that were first down-sampled by a factor of 2, then up-sampling final outputs by $\sqrt{2}$ before performing a second round of pre-training. After these two pre-training rounds, the outputs, $G_x$, $G_k$, and $G_\delta$, are used as initial inputs for the regular training procedure. An 11 x 11 x 11 kernel was used, which was down-sampled to 5 x 5 x 5, and then 7 x 7 x 7 during pre-training stages. The algorithm for updating network weights to predict denoised diffusion images is summarized in Table~\ref{tab:ivim_alg}
    
\begin{table}[h]
            \centering
            \captionsetup{justification=raggedright}
            \caption{The optimization algorithm for updating network weights to predict deblurred PSMA PET images. This algorithm builds off of Ren et al.'s proposed joint optimization algorithm \cite{ren_2020} and implements modifications suggested by Kotera et al. \cite{kotera_2021}.}
            \footnotesize
            \begin{tabular}{@{}l}
            \br
            \multicolumn{1}{c}{\textbf{\large{Diffusion Denoising Algorithm}}}\\
            \br
            \multicolumn{1}{l}{\textbf{Input}: Original 4D (b, z, y, x) image, $\vect{y}$}\\
            \multicolumn{1}{l}{\textbf{Output}: Deblurred 4D (b, z, y, x)image, $\vect{x}$, 4D (b, z, y, x) noise image, $\vect{\delta}$, and 3D blur kernel, $\vect{k}$}\\
            \br
            \textbf{Pre-training}\\
            1. Downsample spatial dimensions of $y$ to 1/2 resolution\\
            2. Initialize $\vect{z_x}$, $\vect{z_\delta}$ from uniform distribution to match size of $\vect{y}$\\
            3. Initialize $\vect{z_k}$ as size $5\times5\times5$ Gaussian kernel with standard deviation of one voxel\\
            4. \textbf{for}\verb| i = 1 to 1000:|\\
            5.     \texttt{~~~~$\vect{x} = G^i_x(\vect{z_x})$}\\
            6.     \texttt{~~~~$\vect{k} = G^i_k(\vect{z_k})$}\\
            7.     \texttt{~~~~$\vect{\delta} = G^i_\delta(\vect{z_\delta})$}\\
            8.     \texttt{~~~~Compute loss and back-propagate gradients}\\
            9.     \texttt{~~~~Update $G^i_x$, $G^i_\delta$ and $G^i_k$ using the ADAM optimizer \cite{adam}}\\
            10.     $\vect{x} = G^{1000}_x(\vect{z_x})$, $\vect{k} = G^{1000}_k(\vect{z_k})$, $\vect{\delta} = G^{1000}_\delta(\vect{z_\delta})$\\
            11. Upsample $\vect{k}$ by 7/5 and upsample spatial dimensions of $\vect{x}$ and $\vect{\delta}$ by a factor of $\sqrt{2}$ \\
            12. Downsample spatial dimensions of $\vect{y}$ by $\sqrt{2}$ to match new convolution size\\
            13. $\vect{z_x} = \vect{x}$, $\vect{z_\delta} = \vect{\delta}$, $\vect{z_k} = \vect{k}$\\
            14. Repeat steps 4 through 10.\\
            15. Upsample $\vect{k}$ by 11/7 and upsample spatial dimensions of $\vect{x}$ and $\vect{\delta}$ by a factor of $\sqrt{2}$ \\
            16. $\vect{z_x} = \vect{x}$, $\vect{z_\delta} = \vect{\delta}$, $\vect{z_k} = \vect{k}$\\
            \br
            \textbf{Main Training} \\
            17. \textbf{for}\verb| i = 1 to 5000:|\\
            18.     \texttt{~~~~$\vect{x} = G^i_x(\vect{z_x})$}\\
            19.     \texttt{~~~~$\vect{k} = G^i_k(\vect{z_k})$}\\
            20.     \texttt{~~~~$\vect{\delta} = G^i_\delta(\vect{z_\delta})$}\\
            21.     \texttt{~~~~Compute loss and back-propagate gradients}\\
            22.     \texttt{~~~~Update $G^i_x$, $G^i_\delta$ and $G^i_k$ using the ADAM optimizer \cite{adam}}\\
            23. $\vect{x} = G^{5000}_x(\vect{z_x})$, $\vect{k} = G^{5000}_k(\vect{z_k})$, $\vect{\delta} = G^{1000}_\delta(\vect{z_\delta})$\\
            \br

        \end{tabular}\\
        \label{tab:ivim_alg}

    \end{table}
        
        The regular training procedure consisted of 5000 iterations, as in previous implementations. This was split into two stages, with two different loss functions. These  stages and their corresponding loss terms are summarized in Table~\ref{tab:loss_func}. It is necessary to include regularization terms in the loss function to avoid convergence to a trivial solution ($\vect{k}$ becomes a unit impulse). Therefore, the mean squared error (MSE) of kernel values above 0.7 were penalized. We only penalized values above this cutoff to avoid a trivial solution while still allowing small contributions from neighbouring voxels to go unpenalized. We furthermore penalized the MSE of noise voxels above $\mu_\delta + \sigma_\delta$, where $\mu_\delta$ and  $\sigma_\delta$ are the mean and standard deviation of the normalized original image voxels located outside of the body. A total variation (TV) loss term for $\vect{x}$ was included after the 3500\textsuperscript{th} iteration, as this has been shown to improve results when employed in later stages of optimization \cite{kotera_2021}. Furthermore, our fidelity loss function switched from MSE to the structural similarity index metric (SSIM) after the 3500\textsuperscript{th} iteration, as recommended \cite{kotera_2021} for improving final image quality.
        
    \begin{table}[h]
                    \centering
                    \captionsetup{justification=raggedright}
                    \caption{Optimization consisted of 5000 iterations, split into 3 stages, each having a different loss function. Up to the 3500\textsuperscript{th} iteration, the fidelity loss function was mean squared error (MSE). The structural similarity index metric (SSIM) was used in remaining iterations. In all iterations, a kernel and noise regularization term were included. The MSE of kernel values were penalized to avoid convergence to the trivial solution. The MSE of noise values above $\mu_\delta + \sigma_\delta$, where $\mu_\delta$ and  $\sigma_\delta$ are the mean and standard deviation of the normalized original image voxels located outside of the body. Finally, the total variation (TV) between denoised image voxels was penalized after the 3500\textsuperscript{th} iteration, as it has been shown to improve results when employed in later stages of optimization \cite{kotera_2021}}
                    \begin{tabular}{@{}lll}
                    \br
                     &Iteration$< 3500$&Iteration $\ge 3500$\\
                    \mr
                    Fidelity Term&MSE& SSIM \\
                    \mr
                     Regularization Terms & 1. Kernel & 1. Kernel \\
                     &2. Noise & 2. Noise\\
                     & & 3. TV \\
                    
                    \br
                \end{tabular}\\
                \label{tab:loss_func}
            
            \end{table}
    \subsection{2.3 Evaluating Denoised Images}
        As neural blind deconvolution is an unsupervised training methodology, ground truth images were not used for direct evaluation of denoised images. We compared image quality between denoised and original IVIM images in terms of two blind image quality metrics. The Blind/Referenceless Image Spatial Quality Evaluator (BRISQUE) \cite{brisque} and Contrastive Language Image Pre-training (CLIP) \cite{clip} score were evaluated. 
        BRISQUE ranks image quality on a scale between 0 and 100 based on a series of features derived from various signal intensity and distribution statistics. These features are used to predict the deviation of the input image from a natural, undistorted image. CLIP uses a language/vision neural network trained using millions of natural image/caption pairs to assess either image quality or caption quality. We furthermore assessed the quality of IVIM parameter maps predicted using original and denoised images in terms of BRISQUE and CLIP metrics.

        We furthermore assessed the similarity between predicted kernels amongst patients. As all patient images were acquired on the same scanner, predicted blur kernels were expected to be similar. As the blur kernel contains components due to both scanner uncertainty, as well as patient motion, we anticipated predicted blur kernels to be similar, yet nonidentical, between patients. We compared blur kernels quantitatively by computing the inner product between normalized kernels. 

        To ensure the model's ability for predicting accurate blur kernels, we employed the same strategy used for evaluating neural blind deconvolution for PSMA PET images \cite{sample_2023_blind_deconv}. This involves generating artificial kernels, convolving them with previously denoised images, then re-starting the neural blind deconvolution process. The objective is then to test how accurately pseudokernels can be predicted, measured using the inner product between generated and predicted pseudokernels. Four types of pseudokernels were generated: a regular Gaussian with a standard deviation of 1 voxel in each direction, and three oblong Gaussians, each having a standard deviation truncated by a factor of three in one of the three Cartesian directions. Pseudokernels were all normalized to a unit sum.

        Original and denoised images were further compared in terms of their practical ability for model-building. Beginning with the hypothesis that IVIM parameters are related to histological features within imaging voxels, we tested the ability of radiotherapy dose levels to predict changes in IVIM parameters following treatment. This was assessed in two different ways. First of all, Spearman's rank correlation coefficient ($r_s$) was compared between the whole-mean dose and relative changes in mean IVIM parameters within each parotid gland ($\frac{P_f - P_i}{{P_i}})$, where $P_i$ and $P_f$ are the parameters before and after radiotherapy). 
        
        The dose response of parameters were further analysed using a multiple regression analysis. Post-radiotherapy parameters were used as dependent variables while pre-radiotherapy parameters and mean dose levels served as two independent variables. The best-fit slope with respect to dose was then compared between parameters. For comparison's sake, slopes were calculated with respect to parameters that had been normalized to statistical Z-values, with statistics derived using pre-radiotherapy voxels within each patient's parotid glands. Images were normalized to statistical Z-value rather than by the maximum image voxel, as it was found that maximum values tended to substantially exceed average values within parotid glands, leading to compression of voxel values. 
        \iffalse
        To make the most of the small dataset, this correlation was also tested in various sub-regions of parotid glands. Parotid glands were divided by planes into 3 equal-volume sub-regions, in each of the three Cartesian patient axes (Figure~\ref{fig:par_subsegs}). The mean dose was correlated with the relative change of mean IVIM parameters in each sub-region.
        \fi
    \subsection{2.4 Extracting Exponential Model Parameters}
        To serve as a basis of comparison for newly derived, model-independent IVIM parameters, we extracted ADC, bi-exponential, and tri-exponential parameter maps for all patients. Voxel-wise parameter fitting was performed using IVIM$_3$-NET models \cite{Kaandorp_2021}, trained for each model type using all acquired original and denoised IVIM images. Voxels were first filtered to exclude voxels located outside the body from training. This resulted in approximately 3.6 million training voxels. Voxels for each b-value were normalized to signals acquired with $b = 0$. IVIM$_3$-NET hyper-parameters were tuned according to previously derived optimal values \cite{Kaandorp_2021}. Models consist of fully-connected networks with 2 hidden layers, which take normalized signal versus b-value arrays as input, and predict the appropriate number of parameters for each model. Each layer consists of a linear function, an exponential linear unit (ELU) activation function \cite{elu}, and batch normalization. Sigmoid functions were used to constrain parameter predictions within boundaries chosen to over-encompass anticipated ranges (Table\ref{tab:param_bounds}). Network weights were then optimized via back-propagation using a batch size of 256 and an ADAM optimizer \cite{adam}.

        To mediate the effect of image noise at $S(b=0)$ affecting all normalized measurements acquired at other b-values, IVIM$_3$-NET models were also used to simultaneously predict $S(b = 0)$ along with other model parameters. This was shown to improve fit results in a previous study \cite{Kaandorp_2021}. Hence, the loss term is of the form,
        \begin{equation}
            L = \left(\frac{S(b)}{\hat{S_0}} - \hat{S}(b)\right)^2
        \end{equation}
        where $S(b)$ is the actual signal in a given voxel, and $\hat{S_0}$ and $\hat{S}(b)$ are model predictions. For example, a bi-exponential model outputs $\hat{S_0}$, $f$, $D^*$ and $D$, which yield $\hat{S}(b) = f e^{-b \cdot D^*} + (1-f)e^{-b \cdot D}$

        %List of tried models / features table
            \definecolor{Dark_Gray}{rgb}{0.5,0.5,0.5}
            \definecolor{Gray}{rgb}{0.7,0.7,0.7}
            \definecolor{whiter}{rgb}{0.8,0.8,0.8}
            \definecolor{white}{rgb}{1,1,1}
            \newcolumntype{a}{>{\columncolor{Gray}}c}
            \newcolumntype{b}{>{\columncolor{whiter}}c}
            \newcolumntype{d}{>{\columncolor{white}}c}
            \begin{table}[h]
                    \centering
                    \captionsetup{justification=raggedright}
                    \caption{Parameter bounds set for IVIM$_3$-NET predictions of exponential model parameters. Bounds were set to over-encompass anticipated parameter ranges, in order to compensate for diminishing gradients at sigmoid asymptotes \cite{Barbieri2019}}
                    \tiny
                    \begin{tabular}{@{}cccc}
            
                    \rowcolor{Dark_Gray}Model&Parameter&Lower Bound&Upper Bound\\
                    \rowcolor{Gray}Bi-exponential&$D$ (mm$^2$/s)&$7\times10^{-5}$&$4\times10^{-3}$\\
                    \rowcolor{Gray}&$D^*$ (mm$^2$/s)&$2\times10^{-3}$&$0.15$\\
                    \rowcolor{Gray}&$f$&$0$&$0.5$\\
                    \rowcolor{white}Tri-exponential&$D$ (mm$^2$/s)&$7\times10^{-5}$&$4 \times10^{-3}$\\
                    \rowcolor{white}&$D^*_1$(mm$^2$/s)&$3\times10^{-3}$&$2\times10^{-2}$\\
                    \rowcolor{white}&$D^*_2$(mm$^2$/s)&$2\times10^{-2}$&0.7\\
                    \rowcolor{white}&$f_1$&0&0.5\\
                    \rowcolor{white}&$f_2$&0&0.5\\
                    \rowcolor{Gray}ADC&$ADC$(mm$^2$/s)&$7\times10^{-5}$&$5\times10^{-3}$\\
                    \br
                \end{tabular}\\
                \label{tab:param_bounds}
            
            \end{table}

    \subsection{2.5 Model-independent IVIM parameters}

        To describe the signal decay curve without imposing specific mathematical models onto the data, we define the area under the curve ($AUC$) parameter. The $AUC$ captures the general fall-off of signal with increasing b-value, and can be equivalently described as the integral of $S(b) / S(0)$. The anticipated relationship between $AUC$ and exponential parameters can be found by integrating over exponential functions between $b = 0$ to $b = 1000$. For the ADC model, 
        \begin{equation}
            AUC = \int^{1000}_0 \frac{S_{ADC}(b)}{S_{ADC}(0)} db = \frac{1}{ADC}\left[1-e^{-ADC \cdot 1000}\right].
        \end{equation}
        $AUC$ and ADC are thus inversely proportional. For the biexponential model,
        \begin{equation}
            AUC = \int^{1000}_0 \frac{S_{bi}(b)}{S_{bi}(0)} db = \frac{f}{D^*}\left[1-e^{-1000 \cdot D^*}\right] + \frac{1-f}{D}\left[1-e^{-1000 \cdot D}\right]
        \end{equation}.
        $AUC$ decreases with increasing D, and is roughly independent of $D^*$ assuming $D^* \approx 10 D$, and f small). For the triexponential model,
        \begin{equation}
            AUC = \int^{1000}_0 \frac{S_{tri}(b)}{S_{tri}(0)} db = \frac{f_1}{D_1^*}\left[1-e^{-1000 \cdot D_1^*}\right] + \frac{f_2}{D_2^*}\left[1-e^{-1000 \cdot D_2^*}\right] + \frac{1-f_1-f_2}{D}\left[1-e^{-1000 \cdot D}\right].
        \end{equation}
        $AUC$ is roughly independent of $D_1^*$ and $D_2^*$ (assuming $D_i^* \approx 10 D$ and $f_i$ small, $i \in \{1,2\}$). 

\begin{figure}[h]
          \centering
          \includegraphics[width=0.9\textwidth]{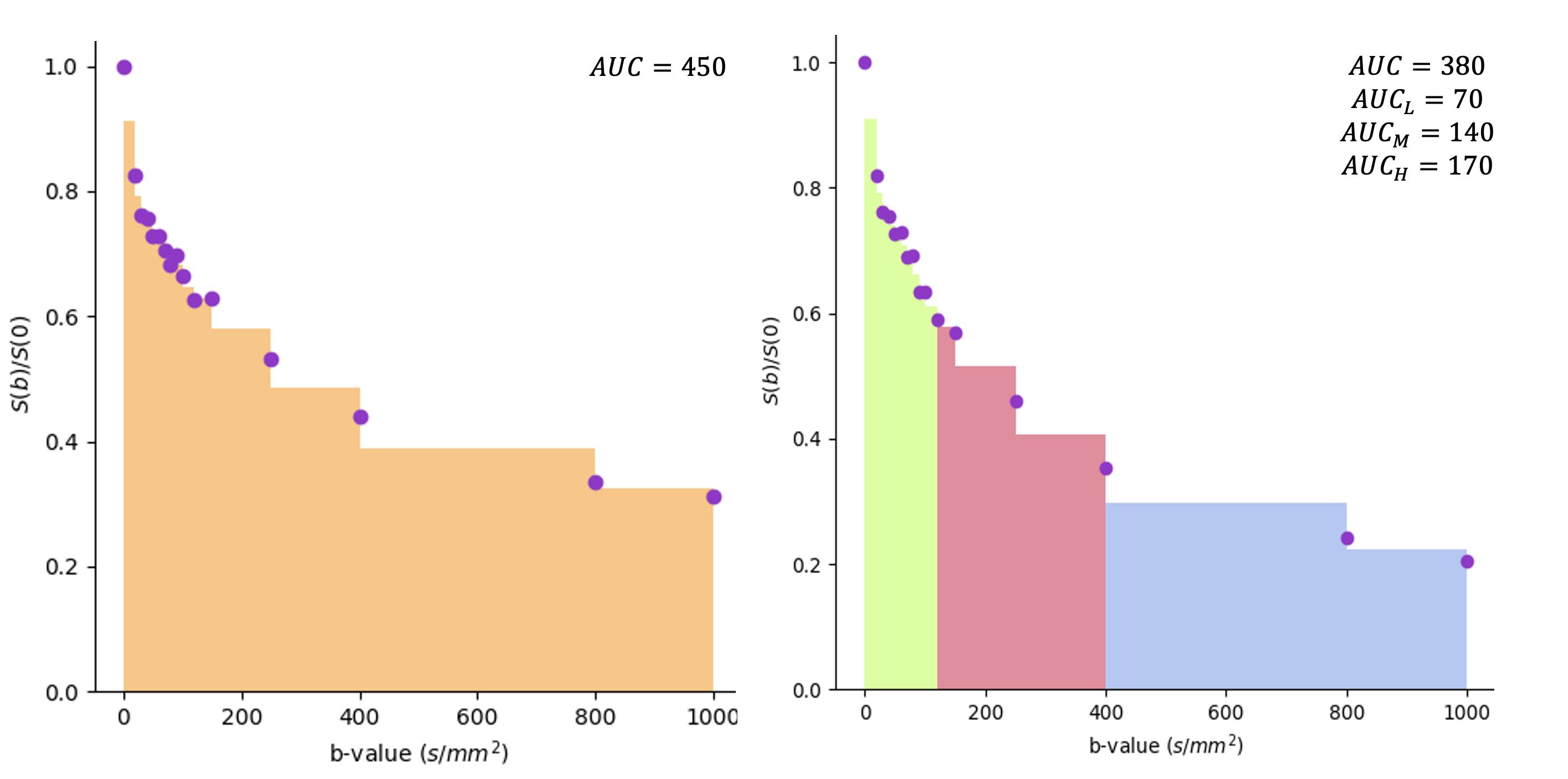}
          \caption{$AUC$ parameters were calculated using middle Riemann sums. On the left, the $AUC$ within a single voxel is calculated graphically. In a separate voxel on the right, $AUC_L$, $AUC_M$, and $AUC_H$ are calculated (yellow, rose, blue), which sum to the regular $AUC$.}
          \label{fig:auc_graph.png}
        \end{figure}

        The $AUC$ was calculated for each voxel using a middle Riemann sum. The AUC was furthermore calculated in 3 subsets of b-values, defining a low-, middle-, and high-range $AUC$ ($AUC_l$, $AUC_m$, $AUC_h$, respectively). The ranges were defined as:
        \begin{itemize}
            \item Low: $b \leq 120$
            \item Medium: $120 \leq b \leq 400$
            \item High: $400 \leq b \leq 1000$
        \end{itemize}
        The method of calculating $AUC$, $AUC_l$, $AUC_m$, and $AUC_h$ using middle Riemann sums is shown graphically in Figure~\ref{fig:auc_graph.png}. 

    \subsection{2.6 Comparing Parameters}
        Whole-mean parameters were computed within all parotid glands. To compare functional utility of $AUC$ parameters with exponential model parameters, we employed several strategies. First of all, we included $AUC$ parameters in the previously mentioned dose-response comparison of IVIM parameters in parotid glands. This allowed the practical utility of parameters for use in dose response studies to be analysed, based on the hypothesis that parameter statistics are related to tissue composition and characteristics at a histological level. 

        We further tested the relative importance of IVIM parameters in capturing the variance of the dataset by evaluating contributions to the primary principal component obtained through singular value decomposition. The primary principal component is the optimal linear combination of all input features for describing the maximum variation in the data. This was computed by first creating a 2-dimensional array, where each row represents a single voxel in an image and each column represents a given feature. Voxels within the body in all cropped images were included as rows in this array. All biexponential, triexponential, $ADC$, and $AUC$ features were included as columns. This yielded an $m\times13$ matrix for which the singular value decomposition was computed. The squared projection of the primary principal component on each of the 13 parameters was then calculated and plotted.
        
    \iffalse
        \begin{figure}[h]
          \centering
          \includegraphics[width=0.8\textwidth]{par_subsegs.png}
          \caption{The ability of radiotherapeutic dose levels to predict changes in IVIM parameters within 9 parotid gland sub-regions was evaluated. Sub-regions were defined by segmenting glands, with parallel planes, into 3 regions of equal volume. This was done with axial (left), coronal (middle), and sagittal (right) planes, yielding a total of 9 sub-regions.}
          \label{fig:par_subsegs}
        \end{figure}
    \fi
\clearpage
\section{Results} 
    Neural blind deconvolution for denoising diffusion MR images appeared to effectively suppress random noise while enhancing fine detail within images. This improvement is supported quantitatively by an increase in blind image quality metrics, BRISQUE and CLIP for denoised images (Table~\ref{tab:biq}). 
    \begin{table}[h]
            \centering
            \captionsetup{justification=raggedright}
            \caption{Blind image quality metrics, CLIP and BRISQUE are compared for original ($\vect{y}$) and denoised ($\vect{x}$) diffusion MR images. Results were averaged over all slices and b-values. The significance of the difference was tested using a paired t-test. }
            \footnotesize
            \begin{tabular}{@{}cccc}
            \br
            &Denoised&Original&Significance\\
            \mr
            BRISQUE&$36.4 \pm 7.8$&$14.6 \pm 6.9$&$p < 0.001$\\
            CLIP&$0.37 \pm 0.03 $&$0.35 \pm 0.03 $&$p < 0.001$\\
            \br
        \end{tabular}\\
        \label{tab:biq}

    \end{table}
    
    Predicted blur kernels were similar among all patients, having a mean inter-patient inner product of $0.88 \pm 0.10$. Furthermore, each patient had two predicted kernels (for diffusion images cropped over the left and right parotid, separately) whose mean inner product was $0.97 \pm 0.02$. Projections of the mean predicted kernel are shown in Fig~\ref{fig:kernel}. Kernel voxels were mainly confined to within axial planes, with the highest variance in the anterior-posterior direction. The general shape of pseudokernels was predicted using neural blind deconvolution; however, it can be seen that predicted kernel voxels were mostly confined to within one voxel of the centre value (Fig~\ref{fig:pseudokernels}).  

    Axial slices of denoised and original diffusion images corresponding to several different b-values are contrasted in Fig~\ref{fig:deblurred_and_orig_v_b}. Image denoising resulted in significantly ($p < 0.001$) more monotonically decreasing signal versus b-value curves, as shown in Fig~\ref{fig:signal_w_decay}. This was assessed by first taking Spearman's rank correlation coefficient, $r_s$, between signal and b-values, averaged over all voxels within body tissue. A paired t-test was then used to assess differences between correlations in denoised and original images. $r_s$ was ${-}0.80 \pm 0.31$ in denoised images and $-0.60 \pm 0.29$ in original images. 

    $ADC$, biexponential, triexponential, and all $AUC$ parameter statistics in parotid glands are summarized in Table~\ref{tab:params}. 

\begin{figure}[h]
          \centering
          \includegraphics[width=0.9\textwidth]{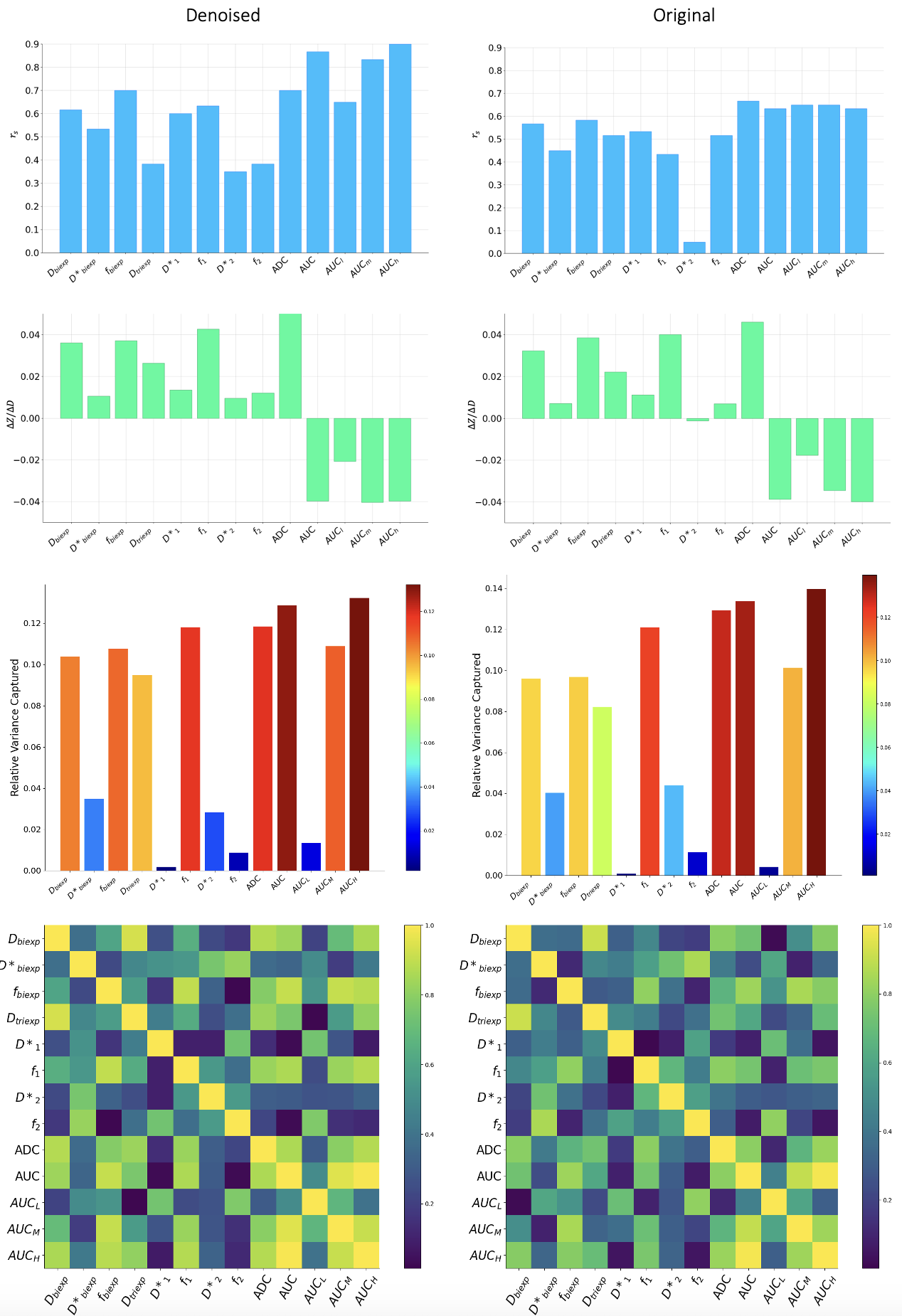}
          \caption{Pairwise Spearman's rank correlation coefficients, $r_s$, between dose and relative changes in biexponential, triexponential, ADC, and AUC parameters derived from IVIM images are displayed (top) for denoised and original images (left and right). Best-fit regression slopes of parameters versus dose are shown underneath in green. The singular value decomposition (SVD) of the data was computed, and the relative variance captured by each parameter in the first principal component was plotted (bottom).}
          \label{fig:svd}
    \end{figure}

%Parameters Table
\definecolor{Gray}{rgb}{0.5,0.5,0.5}
\definecolor{whiter}{rgb}{0.8,0.8,0.8}
\definecolor{white}{rgb}{1,1,1}
\newcolumntype{a}{>{\columncolor{Gray}}c}
\newcolumntype{b}{>{\columncolor{whiter}}c}
\newcolumntype{d}{>{\columncolor{white}}c}
\begin{table}[H]
        \centering
        \captionsetup{justification=raggedright}
         \caption{All IVIM parameter averages and standard deviations inside parotid glands are listed. Averages were calculated over parotid glands from pre-radiotherapy (pre-RT) and post-RT scans separately, along with the significance of their difference. Results are calculated using both denoised and original images. Lastly, the final column lists the significance of differences between denoised and original values before radiotherapy. Significance values were determined using a paired t-test. }
         \tiny
        \begin{tabular}{@{}bcccbbbc}
        \br
        \multicolumn{1}{c}{}&\multicolumn{3}{c}{Denoised Images}&\multicolumn{3}{c}{Original Images}&\multicolumn{1}{c}{}\\
        \rowcolor{white}Parameter&Pre-RT&Post-RT&p-value&Pre-RT&Post-RT&p-value&p-value (Deblur versus Orig)\\
        \mr
        ADC ($\times 10^{-3}mm^2/s$)&$1.23 \pm 0.37$&$1.98 \pm 0.20$& $< 0.001$&$1.27 \pm 0.36$&$1.96 \pm 20$&$<0.001$&$>0.05$\\
        D (biexponential, $\times 10^{-4}mm^2/s$) &$4.8 \pm 1.1$&$6.6 \pm 1.2$&$<0.01$&$4.7 \pm 1.2$&$6.5 \pm 1.2$&$<0.01$&$<0.001$\\
        $D^*$ (biexponential, $\times 10^{-2}mm^2/s$) & $1.6 \pm 0.8$&$1.1 \pm 0.2$&$>0.05$&$2.2 \pm 0.9$&$1.4 \pm 0.4$&$<0.01$&$<0.001$\\
        $f$ (biexponential) & $0.25 \pm 0.04$&$0.32 \pm 0.03$&$<0.001$&$0.26 \pm 0.03$&$0.32 \pm 0.03$&$<0.001$&$<0.001$\\
        $D$ (Triexponential, $\times 10^{-4}mm^2/s$)&$4.6 \pm 1.1$&$6.3 \pm 1.1$&$<0.001$&$4.5 \pm 1.2$&$6.3 \pm 1.1$&$<0.001$& $<0.001$\\
        $D_1^*$ (Triexponential, $\times 10^{-3}mm^2/s$)&$6.2 \pm 0.9$&$5.2 \pm 0.6$ &$>0.05$&$6.9 \pm 1.0$&$6.1 \pm 0.7$&$<0.01$&$<0.001$\\
        $f_1$ (triexponential) & $0.20 \pm 0.04$&$0.27 \pm 0.03$&$<0.001$&$0.20 \pm 0.03$&$0.26 \pm 0.03$&$<0.001$&$p < 0.001$\\
        $D_2^*$ (Triexponential, $\times 10^{-1}mm^2/s$) & $0.70 \pm 0.30$&$0.55 \pm 0.3$&$>0.05$&$1.3 \pm 0.3$&$0.9 \pm 0.3$&$>0.05$&$p < 0.001$\\
        $f_2$ (Triexponential) &$0.04 \pm 0.01$&$0.03 \pm 0.01$&$>0.05$&$0.06 \pm 0.01$&$0.042 \pm 0.001$&$<0.01$&$p < 0.001$\\
        $AUC$ ($s/mm^2$)&$652 \pm 41$&$587 \pm 32$&$<0.001$&$642 \pm 39$&$580 \pm 32$&$<0.001$&$p < 0.001$\\
        $AUC_L$ ($s/mm^2$)&$108 \pm 2.4$&$108 \pm 2.4$&$>0.05$&$105 \pm 1.6$&$107 \pm 1.7$&$>0.05$&$p < 0.001$\\
        $AUC_M$ ($s/mm^2$)&$205 \pm 8.7$&$191 \pm 9.0$&$<0.01$&$201 \pm 7.7$&$189 \pm 7.6$&$<0.01$&$p < 0.001$\\
        $AUC_H$ ($s/mm^2$)&$443 \pm 39.4$&$382 \pm 28$&$<0.01$&$437 \pm 37$&$378 \pm 29$&$<0.01$&$p < 0.001$\\
        \br
    \end{tabular}\\
    \label{tab:params}

\end{table}
Image denoising was found to strengthen correlations between whole-mean parotid dose levels and changes in IVIM parameters (Fig~\ref{fig:svd}). Changes in $AUC$, $AUC_M$, and $AUC_H$ and $ADC$ had the strongest correlations with mean dose levels. Correlations between dose and $AUC$ parameters were particular strong in denoised images. Regression slopes for changes in parameters versus dose are also shown in Fig~\ref{fig:svd}. Changes in $ADC$ had the highest slope with respect to dose in both denoised and original images. $AUC$, $AUC_M$, $AUC_H$, $D^*_2$, $f$, and $D_{biexp}$ had similarly high dose slopes, while $D^*_{biexp}$, $D_{triexp}$, $D^*_1$, $f_1$, $f_2$, and $AUC_L$ had lower slopes. Fig~\ref{fig:svd} also shows the relative contribution of each parameter to the first principal component obtained via the SVD analysis. $AUC$, $AUC_H$ and $ADC$ captured the highest proportion of variance in the data, with $f_1$ also capturing a relatively high proportion of the variance. Pairwise correlations between all parameter combinations are displayed in Fig~\ref{fig:svd}. $AUC$, $AUC_M$, and $AUC_H$ were highly correlated with one another, and also with $ADC$, $D$ (biexponential and triexponential), $f$, and $f_1$, while $AUC_L$ was most correlated with $f_2$ and $D^*_1$. 
    
\begin{figure}[h]
          \centering
          \includegraphics[width=1.1\textwidth]{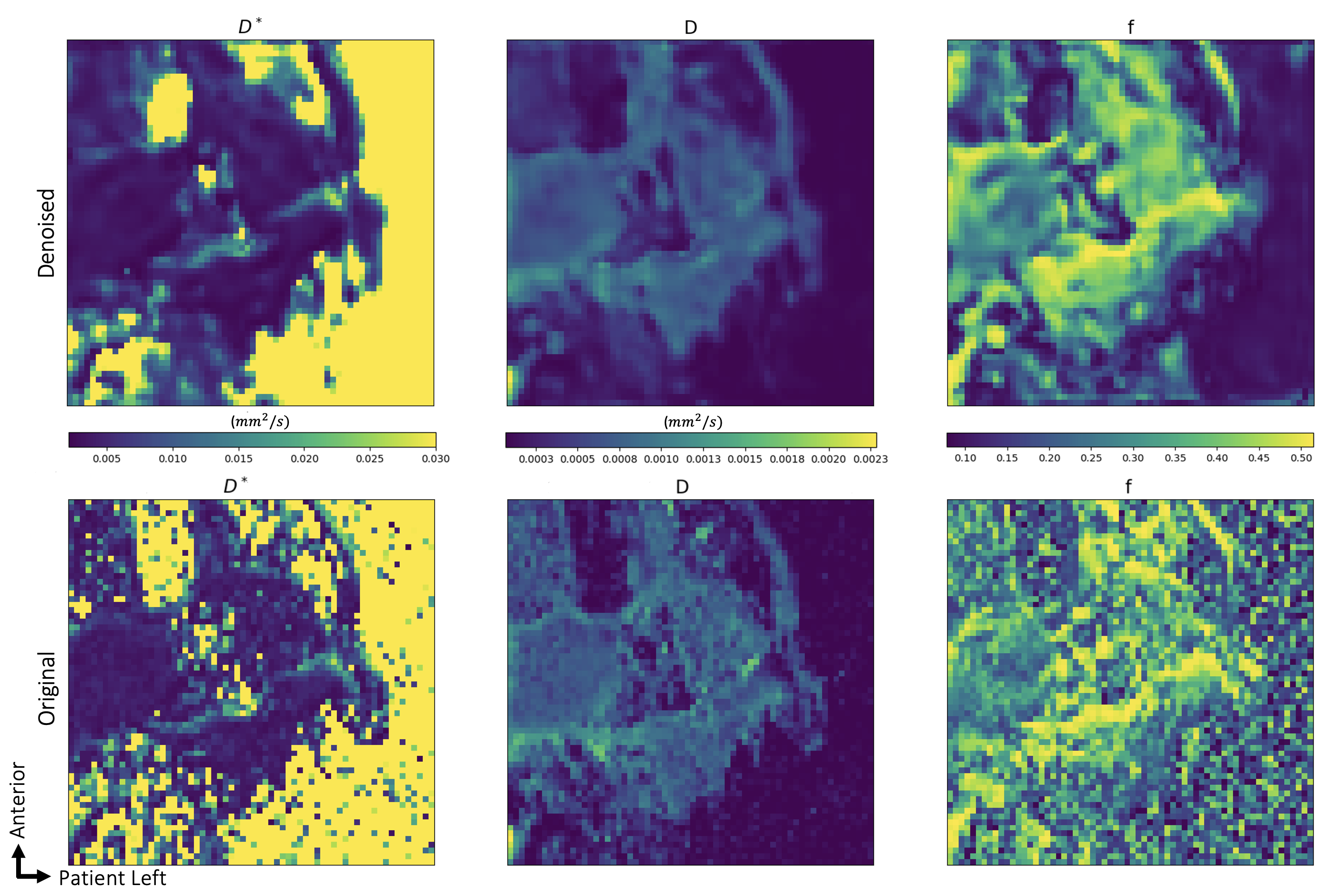}
          \caption{Axial slices through the parotid gland of biexponential model parameter maps are shown, derived using denoised (top) and original (bottom) diffusion MR images. }
          \label{fig:biexp_image_slices}
    \end{figure}

\begin{figure}[h]
          \centering
          \includegraphics[width=1.1\textwidth]{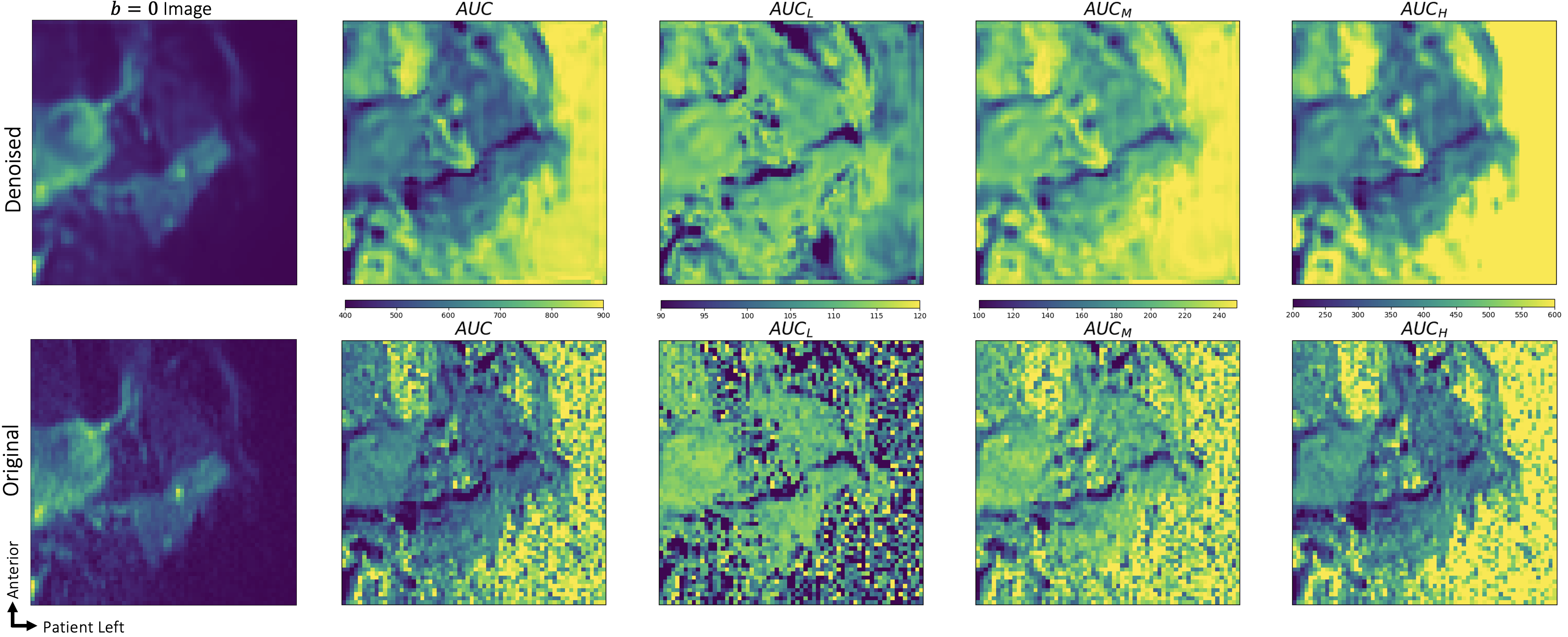}
          \caption{Axial slices through the parotid gland of biexponential parameter maps are shown, derived using denoised (top) and original (bottom) diffusion MR images. Axial $b = 0$, $AUC$, $AUC_L$, $AUC_M$, and $AUC_H$ image slices derived from denoised (top) and original (bottom) are shown.}
          \label{fig:auc_image_slices}
    \end{figure}

     $AUC$ parameter maps derived using denoised and original diffusion images are shown in Fig~\ref{fig:auc_image_slices}. Parameter maps of biexponential parameters corresponding to  the same slice are shown in Fig~\ref{fig:biexp_image_slices}.

    \begin{figure}[h]
          \centering
          \includegraphics[height=1.2\textwidth]{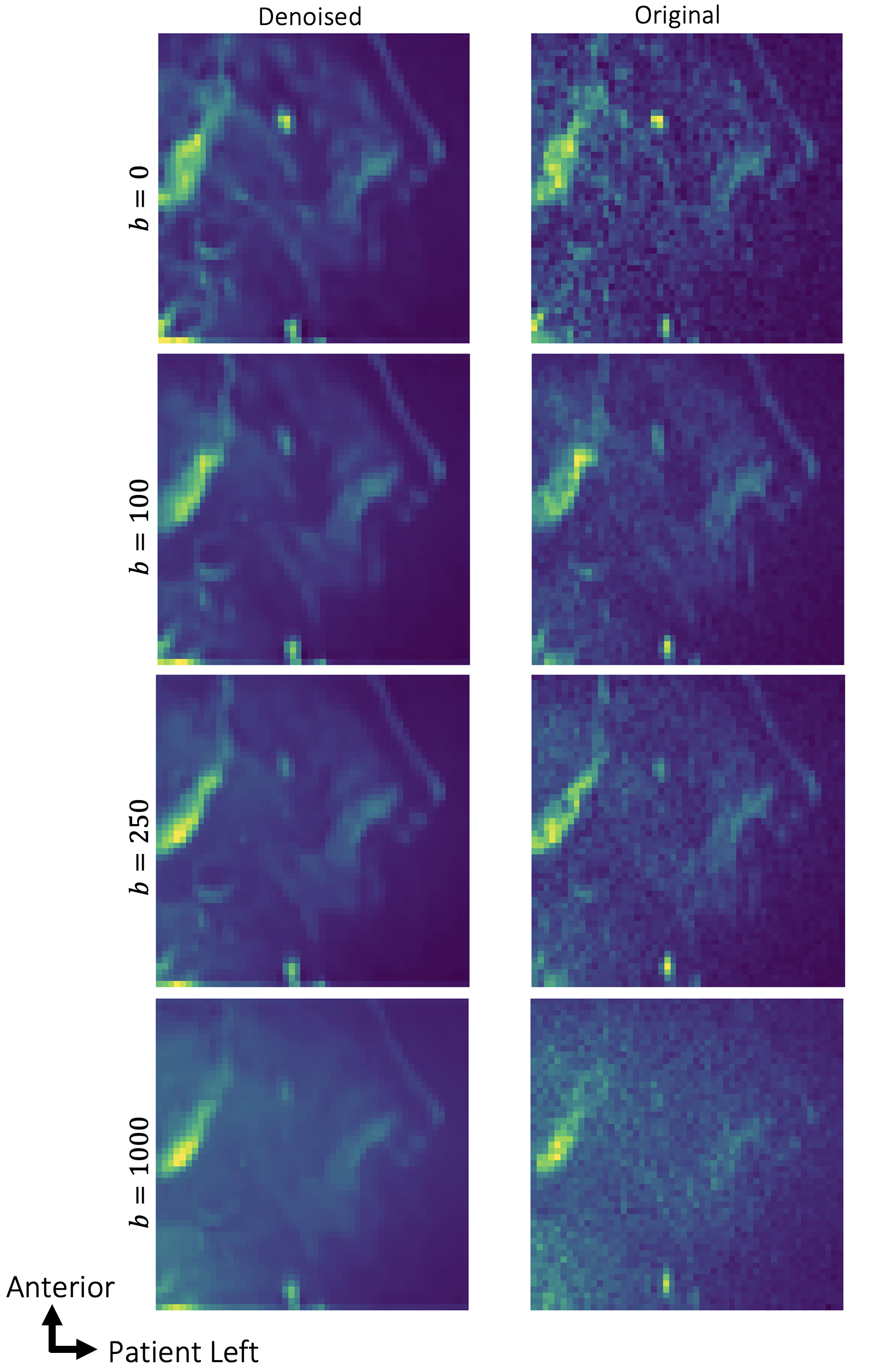}
          \caption{Axial slices through parotid glands of denoised (left) and original (right) diffusion images at 4 different b-values are shown. Denoiding appears to effectively suppress noise while enhancing fine detail within images. Here in particular, the posterior region of the parotid gland becomes discernible only upon denoising.}
          \label{fig:deblurred_and_orig_v_b}
    \end{figure}

    \begin{figure}[h]
          \centering
          \includegraphics[width=1\textwidth]{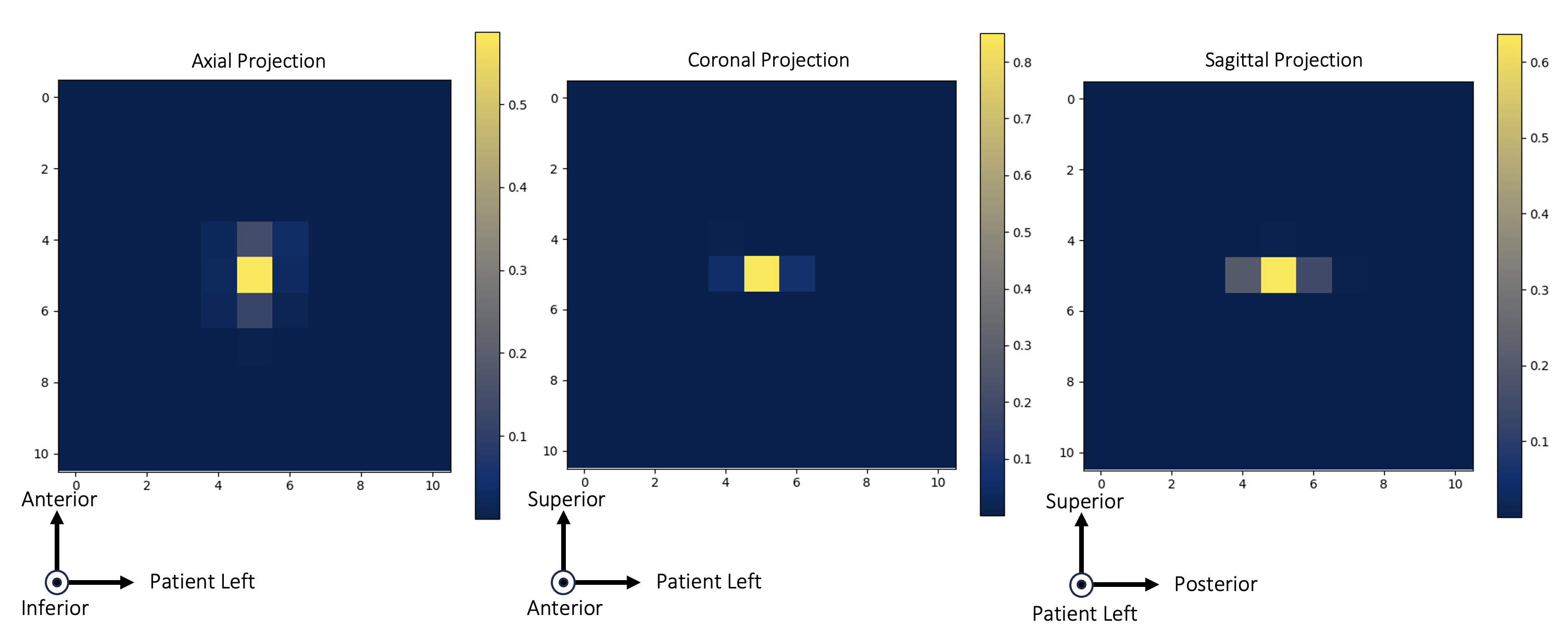}
          \caption{The average blur kernel predicted for patients is projected onto the three primary imaging planes. The kernel was mostly confined to axial planes, which was expected due to elongation of voxels in the superior-inferior direction (slice thickness ${>}2 \times $ axial slice pixel spacing). The kernel had a greater spread in the anterior-posterior direction than the left-right direction.}
          \label{fig:kernel}
    \end{figure}

    \begin{figure}[h]
          \centering
          \includegraphics[width=1\textwidth]{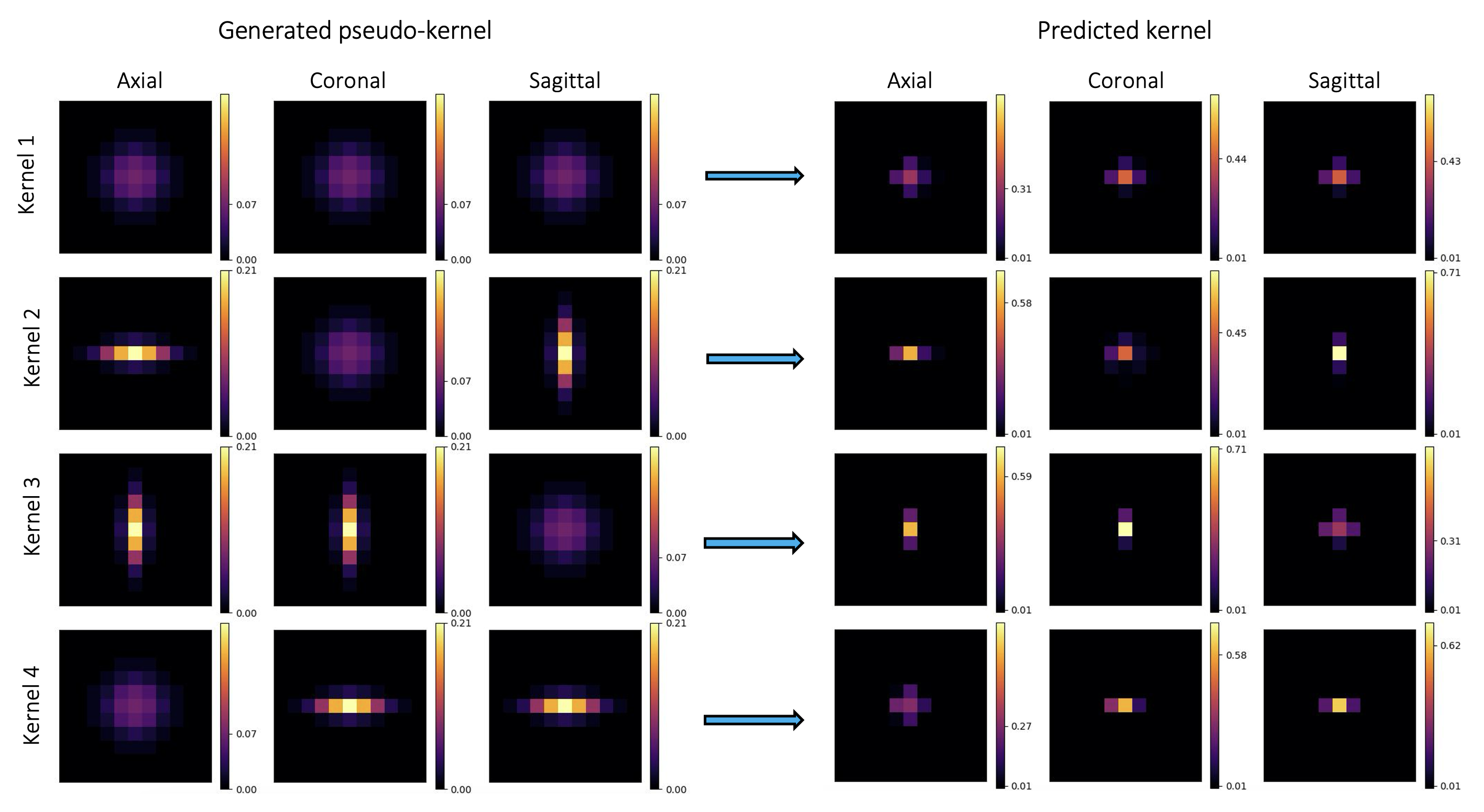}
          \caption{The neural blind deconvolution denoising process was tested by convolving fixed pseudokernels with previously denoised diffusion images before restarting the denoising process. Four pseudokernels, including a regular Gaussian, as well as a Gaussian elongated in each of the 3 primary image axes, were applied. }
          \label{fig:pseudokernels}
    \end{figure}

    \begin{figure}[h]
          \centering
          \includegraphics[width=1.1\textwidth]{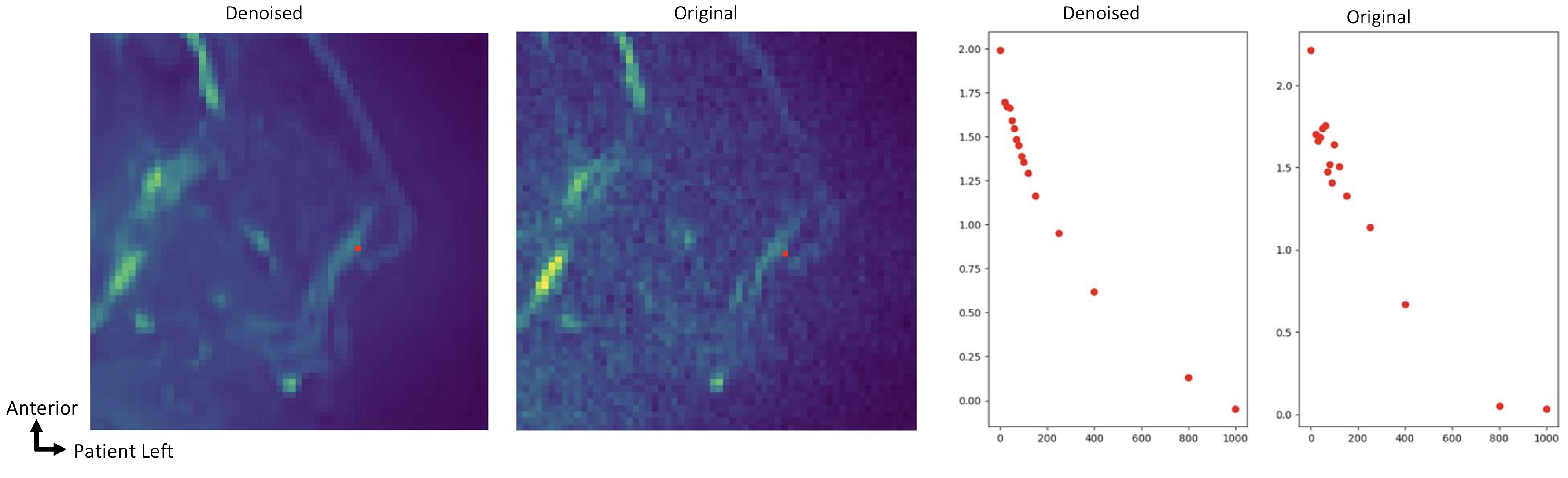}
          \caption{Axial slices through the parotid gland of denoised and original diffusion images with $b = 1000$ are shown, along with the signal decay curve of each in a single voxel (indicated in the denoised image). Signal values were normalized to statistical Z-values within the body. Denoising appears to suppress random noise while illuminating fine detail within images. The loss function used during image denoising did not impose a decrease in signal value with increasing b-value, yet signal values in denoised images appear to naturally decrease more monotonically and more smoothly than in original images.}
          \label{fig:signal_w_decay}
    \end{figure}

\clearpage
\section{Discussion}

There is high variability of derived IVIM MRI parameters in the literature. The proposed model-independent parameter, the $AUC$, proved to be an effective parameter for characterizing the signal decay versus b-value curve. In terms of stand-alone model-building potential, $AUC$ parameters may be more effective than traditional IVIM parameters for predicting various clinical outcomes, based on their high correlations with radiotherapy dose levels and high contributions to the primary principal component of the total dataset. Changes in the $AUC$, $AUC_M$ and $AUC_H$ following radiotherapy were more correlated with dose than diffusion coefficients derived from exponential models. Furthermore, changes in the $ADC$ following radiotherapy were more correlated with dose than diffusion coefficients derived from biexponential and triexponential models. Perfusion fractions for biexponential and triexponential models, $f$ and $f_1$, were more correlated with dose than their corresponding model's diffusion coefficient, $D$.

Neural blind deconvolution for denoising IVIM MRI prior to parameter fitting proved to be effective. Denoising enhanced fine detail in images while suppressing noise, resulting in higher blind image quality metrics, BRISQUE and CLIP \cite{brisque, clip}. denoising resulted in higher correlations between dose and IVIM parameter values, especially with $AUC$, $AUC_M$, and $AUC_H$. This suggests that denoising may be beneficial for future studies seeking to use IVIM imaging to assess radiotherapy-induced tissue damage. $AUC$ values significantly decreased following radiotherapy. $AUC_L$ was found to be the least effective $AUC$ variant, possibly due to the low amount of dataset variability it accounted for, and its lower correlation with dose compared with other $AUC$ variants. This is similar to the limited functional utility of pseudodiffusion coefficients, which also describe the behaviour of the signal decay curve in the low b-value regime. 

In both denoised and original images, the newly proposed Area Under the Curve ($AUC$) of the normalized signal decay curve, and the $AUC$ in the high b-value region ($AUC_H$) captured the largest proportion of the variance. The $ADC$ captured a higher proportion of the variance than the diffusion coefficients of biexponential and triexponential models. Perfusion fractions, $f$ (biexponential) and $f_1$ (triexponential) captured a higher proportion of the variance than diffusion or pseudodiffusion coefficients. Pseudodiffusion coefficients and $AUC_L$ constituted a particularly low proportion of the data's variance. These variables are less correlated with other variables, as seen in the correlations grid.

The results of this study suggest that within our dataset, $AUC$ parameters and the $ADC$ are the most effective IVIM parameters for modelling dose-related outcomes. However, this study included a small number of patients having post-radiotherapy data, rendering our dose response quantities prone to uncertainty. Further validation is needed to assess the efficacy of $AUC$ parameters and neural blind deconvolution for improving the practical utility of IVIM imaging. It appears that denoising images could improve parameter reproducibility, but more work is required for this to be confirmed.

Predicted blur kernels were highly similar between patients, and kernels predicted for images encompassing the left and right parotid gland of each patient were nearly identical (mean inner product: $0.97 \pm 0.02$). Predicted kernels were mainly confined to axial planes, which is expected, due to the axial elongation of the voxel geometry used in this study. To avoid convergence towards a trivial solution, where the kernel becomes a delta function, a regularization term penalizing the MSE of kernel values above 0.7 was included. Previous neural blind deconvolution studies have simply applied MSE to all kernel voxels \cite{kotera_2021, ren_2020}, but we only penalized voxels with values above a given cutoff in order to allow small contributions from neighbouring voxels to go entirely unpenalized. 

This method successfully approximated the general shape of all four pseudokernel types, albeit with truncation at the edges. Predictions were mostly confined to within one voxel of the center value. This indicates that denoising is effective for mitigating cross-talk between neighbouring voxels, but does not account for non-neighbouring voxel interactions. Similar truncation is reported for PSMA PET images \cite{sample_2023_blind_deconv}; however, voxels appear slightly more centralized in this study. This may be partially explained by the smaller kernel size employed in this study. As we were anticipating a relatively small deblur kernel for IVIM MRI images (due to generally sharper MR images than PET), while also considering memory constraints, we decreased the kernel size to 11x11x11 from the 15x15x15 size used for PSMA PET \cite{sample_2023_blind_deconv, sample_2023_psma_importance}. Overall, this denoising process appears to be effective for correcting cross-talk between neighbouring voxels, but more distant effects are not accounted for.

Parameters obtained from denoised and original images had similar population means and standard deviations, but paired t-tests between values found all parameters to be significantly different, except for the $ADC$. $AUC$ values tended to be slightly higher in denoised images than original images, although  differences did not exceed one standard deviation. In all cases, differences in parameters following radiotherapy followed the same trend. Mean ADC values agreed with those previously reported in the literature \cite{Bruvo2021, Nada2017, Kimura2022}. $D$ and $D^*$ were lower than previously reported values \cite{Zhou2016, Becker2016, Kimura2022}, both before and after radiotherapy. However, $D$ and $D^*$ were within a standard deviation of the mean values reported by Kimura et al. \cite{Kimura2022}. Furthermore, diffusion coefficients increased after radiotherapy with similar proportions to those found previously \cite{Zhou2016}. $f$ within parotid glands was lower than values reported by Zhou et al. \cite{Zhou2016} and Beckert et al. \cite{Becker2016} but smaller than values reported by Kimura et al. \cite{Kimura2022}. 

It is not uncommon for IVIM parameters reported in the literature to be in conflict \cite{Lemke2011}, which was the motivation for this work. For example, the significant variability of reported IVIM parameters in the liver has created difficulties establishing normal, baseline perfusion parameters \cite{Cieszanowski2018}. The low SNR associated with echo-planar imaging is likely a significant contributor to these issues. Disregarding time constraints, a spin echo sequence may lead to much greater reliability in parameter estimates. Variations in SNR between scanners used in different studies has been reported to affect parameter estimates \cite{Cieszanowski2018, Liao2021}. Variability of b-value distributions \cite{Dyvorne2014} and voxel sizes also contribute to variability among parameter estimates in the literature.

The intrinsically low SNR of echo planar imaging sequences used for IVIM imaging makes denoising an important endeavour. Incorporating a noise term into the neural blind deconvolution methodology allowed for corrections of non-negative, scanner related noise to be corrected for. Noise suppression after denoising is clearly visible in voxels located outside of body regions (Fig~\ref{fig:deblurred_and_orig_v_b}). A notable result is the increased monotonicity of the signal decay curve found after image denoising (Fig~\ref{fig:signal_w_decay}). It is expected that increasing diffusion gradient strength should result in strictly decreasing signal acquisition; however, this expectation was not enforced in the loss function used for optimization. This perceivable suppression of noise in the decay curve seems to validate the denoising process. 

Characterizing the signal decay curve non-parametrically can be advantageous, as it has the potential to improve the utility and reproducibility of IVIM parameters without presuming the underlying mathematical model. While it is more desirable in theory to have physically interpretable parameters, it has not proven to be practical for IVIM imaging. The AUC and its sub-quantities are only one method of describing the signal decay curve. While this study affirms their utility, there remains room for further exploration and validation of other model-independent parameters. Future studies can continue to improve upon this methodology using the code available online \cite{sample_git}.

\section{Conclusion}
Characterizing the signal decay curve using model-independent metrics, such as the the $AUC$ parameters, and denoising images prior to parameter estimation, could improve reproducibility and the practical utility of IVIM imaging for developing outcome-predictive models. IVIM parameters have shown substantial variability throughout the literature. A shift in the methodology for characterizing the signal decay curve offers a potential solution to mitigate this inconsistency.

\section{Acknowledgements}
We thank the entire MRI staff, and radiation oncologists at BC Cancer Vancouver for helping with the data collection. We thank Mrs. Rosalie Segal and the late Mr. Joseph Segal for funding this research. We acknowledge an in-kind software loan from Siemens Healthcare Limited.

\section*{References}

\bibliography{bib} 
\bibliographystyle{ieeetr}

\end{document}